# Micro Embossing of Ceramic Green Substrates for Micro Devices

Xuechuan Shan[*1], S. H. Ling[1], H. P. Maw[1], C. W. Lu[1] and Y. C. Lam[2]

[1]Singapore Institute of Manufacturing Technology (SIMTech), 71 Nanyang Drive, Singapore 638075

[2]Mechanical and Aerospace Engineering, Nanyang Technological University, Singapore 639798

[+] xcshan@simtech.a-star.edu.sg (Shan); Tel: +65-67938560

*Abstract*- **Multilayered ceramic substrates with embedded micro patterns are becoming increasingly important, for example, in harsh environment electronics and microfluidic devices. Fabrication of these embedded micro patterns, such as micro channels, cavities and vias, is a challenge. This study focuses on the process of patterning micro features on ceramic green substrates using micro embossing. A ceramic green tape that possessed near-zero shrinkage in the x-y plane was used, six layers of which were laminated as the embossing substrate. The process parameters that impact on the pattern fidelity were investigated and optimized in this study. Micro features with line-width as small as several micrometers were formed on the ceramic green substrates. The dynamic thermo-mechanical analysis indicated that extending the holding time at certain temperature range would harden the green substrates with little effect on improving the embossing fidelity. Ceramic substrates with embossed micro patterns were obtained after co-firing. The embedded micro channels were also obtained by laminating the green tapes on the embossed substrates.**

## I. Introduction

Multilayered ceramic substrates with embedded micro patterns are becoming increasingly important, for example, in harsh environment electronics and microfluidic devices [1]. These micro patterns, which include micro channels, cavities and vias, as well as an assembly of them, are the constituent parts for microfluidic devices; they can also be filled with conductive or dielectric materials to form high aspect interconnections or passive components for multi-functional substrates. The embedded micro patterns are conventionally fabricated by mechanical punching or laser machining on ceramic green sheets, followed by lamination and sintering [2-3]. These conventional fabrication methods, however, have unavoidable limitations, for instance, the minimum features are limited by the size of the machining tools, and the minimum depth is restricted by the layer thickness of a ceramic green tape.

Micro hot embossing is a cost-effective technique for manufacturing micro structures on polymeric materials [4]. The mechanisms for cavity filling and process optimizations in polymer embossing have been widely investigated and reported by many research groups [5-7], and it has been successfully used for manufacturing micro components for fluidic devices and systems.

We proposed to utilize micro embossing for patterning micro structures on ceramic green substrates, followed by lamination and sintering to manufacture embedded micro patterns [8]. In this paper, we present our study on the process issues of embossing ceramic green substrates. A ceramic green substrate composes of ceramic-based powders and polymeric additives; the proportion of polymeric additives is approximately 20 wt%. This makes the embossing of ceramic green substrates different from polymer embossing. The challenges of embossing a ceramic green substrate include the following aspects: (1) material flow during embossing is slow; (2) the embossed substrate will be torn or deformed during demolding due to its low strength. To address these fundamental process issues in ceramic embossing is necessary and important.

In our study, a ceramic green tape that possesses near-zero shrinkage in x-y plane was employed, six layers of which were laminated for micro embossing. The dynamic thermo-mechanical analysis showed that extending the holding time at a certain temperature range would harden the green substrates; this would increase the difficulty in achieving high quality embossing. The process parameters that impact on the embossing fidelity were investigated. Micro patterns with a feature size of several micrometers were formed on ceramic green substrates. The embedded micro channels were obtained by laminating the embossed substrate with green tapes, followed by sintering.

## II. Experimental Designs

### A. Micro Embosser and Embossing Mold

A desk-top hot embosser was used, which utilized a hydraulic pressing mechanism for supplying pressure with a maximum working force of 150,000 kN. It possessed two pressing platens - the top and bottom platens; each of which was 100 mm in diameter and equipped with an embedded heater. Both the top and bottom heaters could be preset to a maximum temperature of 250 °C with a tolerance of ± 2°C. After embossing, cooling was achieved by means of water-cooling blocks attached to the platen as well as natural air convection. A nickel template with micrometer-scaled cavities and channels was employed as the mold for embossing. The mold was replicated from a silicon master via electroplating; the plated mold was about 1 mm thick and was coated with a thin anti-sticking layer (100 nm thick) to improve the ease in demolding.

*B.  Preparation of Ceramic Green Substrate*

Most of the ceramic green tapes will suffer shrinkage in all directions upon sintering; the tolerance control of the shrinkage in each direction will affect the final dimension and location of the components and features fabricated on the tapes. A ceramic green tape, the HeraLock™ HL2000 from Heraeus, was used, which possessed near-zero shrinkage in the x-y plane, densifying itself by shrinking primarily in the *z* direction. This shrinkage property would simplify designs in feature dimension and component layout. The thickness of a single layer tape was 127 μm. Six layers of the green tapes were precisely aligned and then laminated to form a stack, which was used as the substrate for micro embossing. The lamination was carried out by using an isostatic thermal laminator with a pressure of 100 kgf/cm$^2$ (9.8 MPa) at 75 °C for 5 minutes.

*C.  Experimental Methodology*

The laminated green substrate was characterized by means of a thermo-gravimetric analyzer (TGA) to determine the weight ratio of polymeric additives; its thermal properties were characterized via dynamic mechanical analyzers (DMA) to determine its glass transition temperature to facilitate the choice of suitable process parameters. Three groups of embossed patterns with regular channel-width (S) and protrusive line-width (L) were targeted.

(1) Group A: the channel-width S and line-width (L) were 50 μm and 100 μm (S/L = 0.5), respectively.

(2) Group B: S and L were 100 μm and 100 μm respectively, and S/L = 1.0;

(3) Group C: S and L were 100 μm and 400 μm respectively, and S/L = 0.25;

The profile of the embossed patterns was measured and observed by means of a profilometer and a microscope.

### III.  EXPERIMENTAL RESULTS

*A.  Characterizations of Ceramic Green Substrates*

Thermo-gravimetric Analysis (TGA), which measures the changes in weight of a sample as a function of temperature, was performed with the purpose of determining the percentage and thermal degradation temperature of organic additives in the green substrate. TGA was carried out with air as the purging gas and the sample size was approximately 8.0 mg. Fig. 1 shows the percentage of weight change of the ceramic green substrate with temperature ranges from 30°C to 800°C at a heating rate of 1°C/min, 5°C/min, 10°C/min and 20°C/min. For all the four different heating rates illustrated in Fig.1, the thermal degradation of the organic additives showed two obvious regions, although the initiation and completion temperatures of thermal degradation were different for different heating rates. Taking the heating rate of 1°C/min as an example, the TGA curve shows that the green substrate lost about 0.4 wt% when the temperature reached 110 °C; this was attributable to the evaporation of plasticizer. The ceramic green substrate continued to lose its weight as temperature increases and it lost about 16.5 % of its weight when the temperature reached 350 °C. There was not much weight change thereafter up to 800 °C. This indicated that this particular ceramic green tape consisted of about 16.5 wt% of organic additives and 83.5 wt% of ceramic particles.

The ceramic green tape used in this study was flexible at room temperature, but its thermal behaviour at raised temperature was unknown. Similar to polymer embossing, the glass transition temperature ($T_g$) of the ceramic green material was an important indicator for determining its embossing temperature. Therefore, the dynamic mechanical analysis (DMA) was conducted to determine both its viscoelastic and thermal properties.

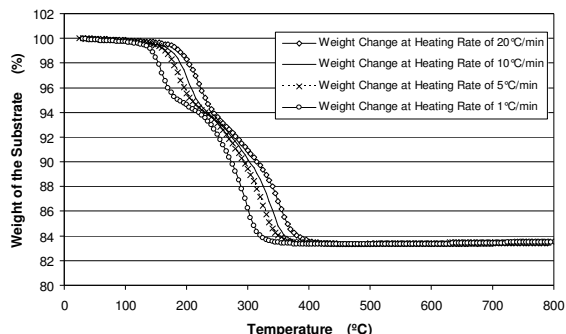

Fig.1. TGA curves of green substrate at different heating rates. The weight loss below 110 °C was attributable to evaporation, and the thermal degradation below 400 °C showed two obvious regions.

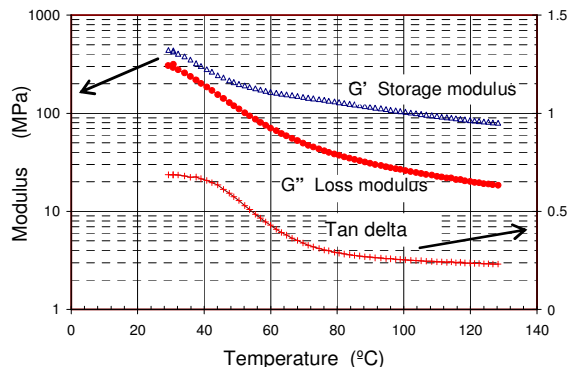

Fig. 2. DMA curves for analyzing glass transition temperature of ceramic green substrate.

Fig.2 illustrates the results obtained from dynamic mechanical analysis (DMA Q800). The sample was a six-layer laminate, which was the same as that used for embossing. The temperature varied from room temperature to 130 ºC, with the temperature ramp and frequency being 2ºC /min and 1 Hz, respectively. Both the storage modulus (G′) and loss modulus (G″) decreased as the temperature increased to 130 ºC. The peak of tan δ (tan δ= G″/G′) indicated that the glass transition temperature ($T_g$) of the green substrate ranged from 35-50 ºC. The increase of temperature above its $T_g$ resulted in further decrease of moduli G' and G″. As a result, the green substrate softened and became more flexible with an increase in temperature, which is advantageous for micro embossing. It was also found that the moduli G″ and G' would not decrease significantly while the temperature was further increased above 60 ºC. Hence, a good embossing temperature would range from 60 ºC to 100 ºC.

### B. Effect of Applied Force

The ceramic green tapes consist of solid powders and organic additives. The organic additives are less than 16.5 wt%; this makes the embossing of green substrates more difficult since a lower ratio of polymeric content will limit and slow down the material flow in embossing. Nevertheless, micro embossing of ceramic green substrates was successfully performed using the six-layered lamination. Fig. 3(a) demonstrates the embossed result of a tapered channel; its sharp end was less than 5 μm. This result indicated that micrometer-scaled features can be formed on ceramic green substrate using embossing. Fig. 3(b) shows cross section of the embossed micro channel array with the ratio of channel-width/line-width being 100μm/100μm. The sidewalls of the channels were slightly collapsed on dicing the green substrate for showing the cross section. Hence, the actual fidelity would be better than illustrated in Fig. 3(b).

The impact of the applied forces to the embossed pattern fidelity was investigated when the embossing temperature and holding time were 70 °C and 3 minutes, respectively. Fig. 4 illustrates the embossed depth for group B (channel-width/line-width = 100μm/ 100μm) when the applied force changed from 10 kN (equivalent to an average pressure of 2.5 MPa) to 20 kN and 25 kN. Fig. 5 shows the effect of the force on the embossed depth of the patterns in Group B and Group C (channel-width/line-width = 100μm/ 400μm). Fig. 5 indicates that the embossed depth depended not only on the applied force but also on the pattern design or pattern density. For the targeted two groups of patterns, it was estimated that a applied force of 20 kN or above was suitable to achieved the required embossing depth.

### C. Effect of Embossing Temperature

The impact of embossing temperature was investigated with the applied force kept at 20 kN (equivalent to an

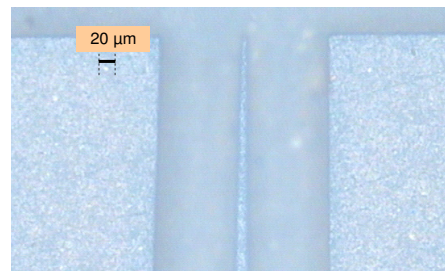

(a) An embossed tapered channel

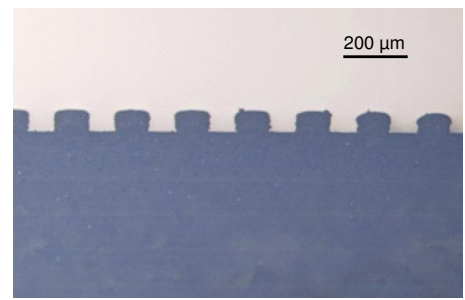

(b) Cross section of the embossed patterns with channel-width/ line-width being 100μm /100μm.

Fig. 3. Patterns embossed on ceramic green substrates

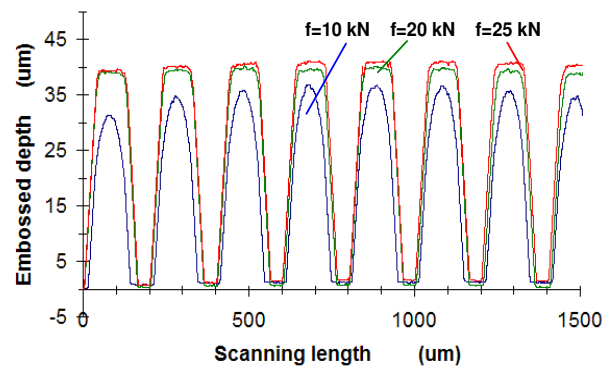

Fig.4. The effect of applied force on embossed depth of Group B patterns with channel width /line width being100 μm/ 100μm.

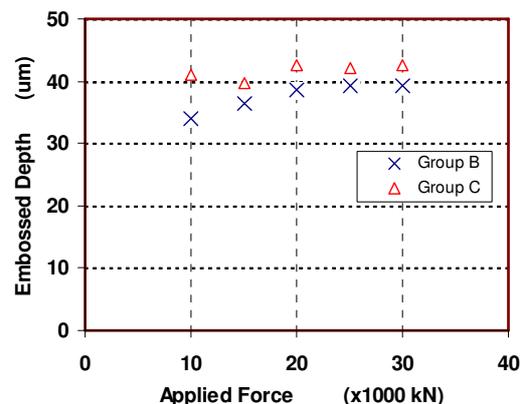

Fig.5. The effect of applied force on embossed depth. Group B and Group C represent the channel-width/line-width of 100μm/100μm and 100μm/400μm, respectively.

average pressure of 5.0 MPa) for 3 minutes before demold. Fig. 6 illustrates the measured depth of the embossed channels in Group A (channel-width/line-width = 50μm /100μm) under different temperatures. The embossed depth increased with an increase in process temperature, and the increase in depth was obvious when the temperature was below 70 °C. However, there was no significant increase when the temperature was above 70 °C. Embossing above 120 °C made it difficult to separate the substrate from the mold; in some cases the embossed patterns were elongated during demolding. Under the same temperature and applied pressure, lower pattern density (lower ratio of channel-width/line-width) resulted in larger embossed depth of channels. Fig. 7 indicated that the embossed depth of Group C (channel-width/ line-width = 100μm/ 400μm) reached the maximum depth even at 50 °C.

Fig. 8 summarized the influence of the embossing temperature to embossing depth. Conclusion could be made for the three targeted groups that stable embossed depths were achieved when the embossing temperature was set above 70 °C.

*D. The Effect of Time Factor*

The effect of holding time at peak applied-force was investigated. For the three targeted groups of patterns, the effect of holding time was not obvious. It appears that 3 minutes was a suitable and sufficient period of holding time to get the maximum embossed depth. In the case that the pattern was not completely replicated within 3 minutes, extending the holding time increased the depth of macro deformation [8] but demonstrated little improvement in embossed depth of tiny patterns, and this was true especially for high-density micro patterns. This is due to the low fluidity of the ceramic green material to fill up the micro patterns especially in the vertical direction.

The force ramp could be defined as the rate of force increment with respect to time. In the cases of 10 kN or 20 kN, the force was increased to its peak value in 5 seconds, 10 minutes and 20 minutes. It was found that a slow force ramp reduced the macro deformation in a certain degree but showed little effect in improving the embossed depth of micro patterns.

Thus, the experimental results indicate that both the holding time and force ramp showed little effect for improving the fidelity of micro pattern formation. This rather "time independent" behaviour of the ceramic green substrate in micro embossing is likely due to the following reasons: (1) The low viscoelasticity and low fluidity of the green material, which were attributed by the amount of polymeric additives in the material, slowed down and/or limited the filling of tiny patterns, thus resulted in incomplete embossing. (2) The little material flow contributed in filling the porosities in green substrate rather than pattern formation. (3) The thermal properties of the green substrate had changed during the embossing. Further investigations were required to identify the key reasons and solutions for improving embossing fidelity.

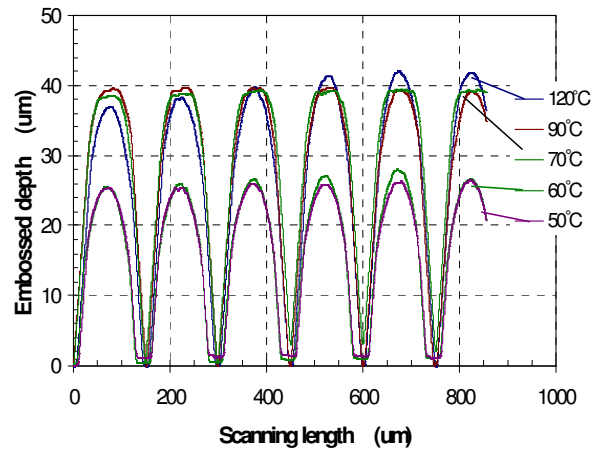

Fig.6. Effect of embossing temperature on patterns in Group A with channel-width/line-width being 50μm/100μm.

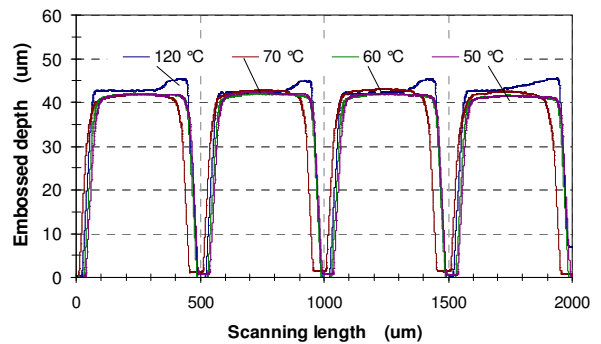

Fig.7. Effect of embossing temperature on patterns in Group C with channel-width/line-width being 100μm/400μm.

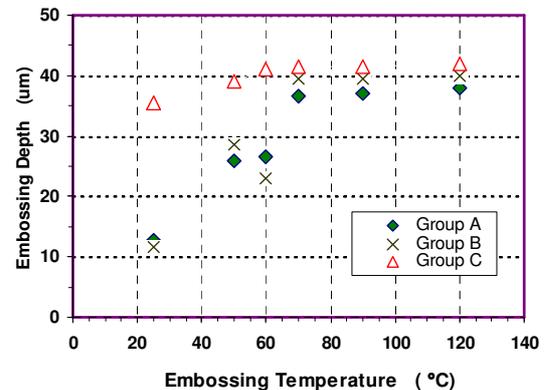

Fig.8. Embossing temperature versus embossed depth of patterns with different pattern density. The density ratios (channel-width/line-width) of Groups A, B and C were 0.5, 1 and 0.25, respectively.

IV. DISCUSSIONS

To investigate both the viscoelastic and thermal properties at raised temperature of the green substrate, dynamic mechanical thermal analysis (DMTA) in torsion was conducted by using a rotational rheometer – Physica

MCR from Anton Paar with air as the purging gas. The specimens with a rectangular shape of 36 mm× 10mm× 0.7 mm were prepared, which were subjected to a tensile force of 0.2±0.05N (frequency: 1Hz and percentage strain γ: 0.01%) with a heating rate of 1ºC/min.

Fig. 9 presents the viscoelasticity behaviors of the green substrate at isothermal condition with holding time of 20 minutes. These tests were conducted with an open hysteresis loop by heating up the specimens at room temperature to the desired process temperature at 1ºC/min, hold the specimens isothermally for 20 minutes, and then cooling down to 30ºC. It was found that both the storage modulus (G′) and the loss modulus (G″) were decreasing with an increase in temperature, which indicated that the green substrate was being softened. At process temperature below 80 ºC, there was a negligible change in G′ and G″ during the holding period. However, when the process temperature increased to or above 100ºC, the value of storage modulus G′ increased with holding time whereas G″ decreased. This indicated that the thermodynamic property of the green substrates had altered at these temperature ranges. The organic additives were believed to have partially decomposed. This led to the increases of G′ with holding time. The increase of G′ implies that the substrate behaves more elastically. The decrease of G″ could mean that the heat supply was not enough to soften the second type of organic additives in the green substrate. In any case, it could be predicted that the embossing process carried out at such temperature would result in poor embossing fidelity, and the fidelity would not be significantly improved by increasing the holding time at peak temperature. These observations are consistent with our embossing results.

V. CONCLUSIONS

Micro embossing of ceramic green substrates was demonstrated. The process parameters that impact on the embossing fidelity were studied. Micro features with several micrometers and micro channels with channel-width as small as 50 µm were formed. The quality and fidelity of embossed patterns depended not only on the applied force (pressure) and temperature, but also on the pattern density. The influence of holding time and force ramp were more complicated; simply extending the holding time or using low force ramp demonstrated little effect in improving the embossed depth or embossing quality. Nevertheless, the achievements of this study demonstrated that micro embossing is a promising technique for fabricating microstructures on ceramic green substrates.

ACKNOWLEDGMENT

The authors would like to thank Mr. Ricky T. Rjeung for his contribution. This work is supported by Agency for Science, Technology and Research (A*STAR) of Singapore for Singapore-Poland Cooperation.

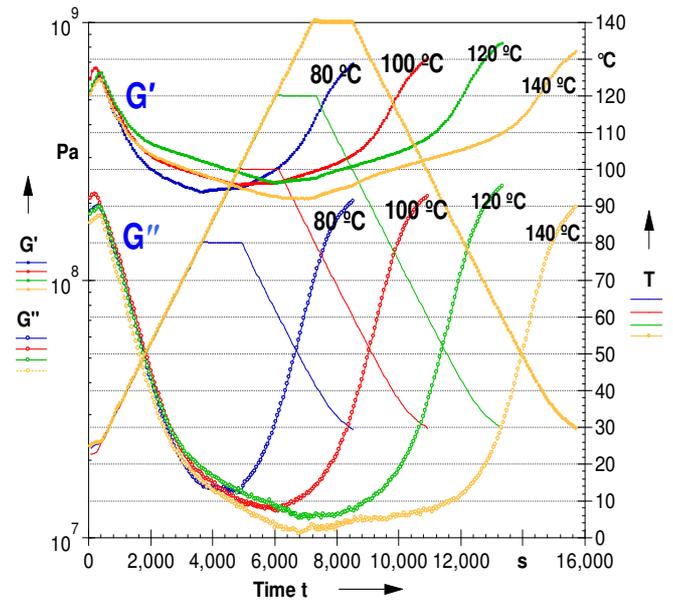

Fig. 9. Effect on storage modulus G′ and loss modulus **G″** at different temperatures (80 ºC, 100 ºC, 120ºC and 140ºC) with a holding time of 20 minutes